\documentclass[a4paper,11pt]{article}
\usepackage{subfigure}
\usepackage{graphicx}
\usepackage{pos}
\usepackage{lineno}
\usepackage{xspace}
\usepackage{lineno}
\usepackage{color}

\title{$\psi(2S)$ production as a function of charged-particle multiplicity in pp collisions at $\sqrt{s}$= 13 TeV and p--Pb collisions at $\snn$ = 8.16 TeV with ALICE at the LHC}
\ShortTitle{$\psi(2S)$ production  as a function of charged-particle multiplicity}

\author*{Theraa TORK}
\author{on behalf of the ALICE collaboration}

\affiliation[a]{Laboratoire de Physique des 2 Infinis, Irène Joliot-Curie,\\
 Orsay, France}

\emailAdd{theraa.tork@cern.ch}

\abstract
{ 
    
   The self-normalised $\psip$ yields as a function of charged-particle multiplicity in pp collisions at $\s$ = 13 TeV and p--Pb collisions at $\snn$ = 8.16 TeV, using the ALICE detector at the  CERN LHC, are reported in these proceedings. This measurement represents one of main tools to understand the role of multiparton interactions and their influence on charmonium production in small systems, i.e pp and pA collisions. The ratio of $\psip$-over-J/$\psi$ as a function of charged-particle multiplicity is also presented for both pp and p--Pb collisions. The comparison to theoretical calculations is reported.     

}

\FullConference{
  The Tenth Annual Conference on Large Hadron Collider Physics - LHCP2022\\
  16-20 May 2022\\
  online
}

\begin{document}
%

\newcommand{\pp}           {pp\xspace}
\newcommand{\ppbar}        {\mbox{$\mathrm {p\overline{p}}$}\xspace}
\newcommand{\XeXe}         {\mbox{Xe--Xe}\xspace}
\newcommand{\PbPb}         {\mbox{Pb--Pb}\xspace}
\newcommand{\pA}           {\mbox{pA}\xspace}
\newcommand{\pPb}          {\mbox{p--Pb}\xspace}
\newcommand{\Pbp}          {\mbox{Pb--p}\xspace}
\newcommand{\AuAu}         {\mbox{Au--Au}\xspace}
\newcommand{\dAu}          {\mbox{d--Au}\xspace}

\newcommand{\s}            {\ensuremath{\sqrt{s}}}
\newcommand{\snn}          {\ensuremath{\sqrt{s_{\mathrm{NN}}}}\xspace}
\newcommand{\pt}           {\ensuremath{p_{\rm T}}\xspace}
\newcommand{\meanpt}       {$\langle p_{\mathrm{T}}\rangle$\xspace}
\newcommand{\ycms}         {\ensuremath{y_{\rm cms}}\xspace}
\newcommand{\ylab}         {\ensuremath{y_{\rm lab}}\xspace}
\newcommand{\etacms}       {\ensuremath{\eta_{\rm cms}}\xspace}
\newcommand{\etalab}       {\ensuremath{\eta_{\rm lab}}\xspace}
\newcommand{\etarange}[1]  {\mbox{$\left | \eta \right |~<~#1$}}
\newcommand{\yrange}[1]    {\mbox{$\left | y \right |~<~#1$}}
\newcommand{\dndy}         {\ensuremath{\mathrm{d}N_\mathrm{ch}/\mathrm{d}y}\xspace}
\newcommand{\dndeta}       {\ensuremath{\mathrm{d}N_\mathrm{ch}/\mathrm{d}\eta}\xspace}
\newcommand{\avdndeta}     {\ensuremath{\langle\dndeta\rangle}\xspace}
\newcommand{\dNdy}         {\ensuremath{\mathrm{d}N_\mathrm{ch}/\mathrm{d}y}\xspace}
\newcommand{\Npart}        {\ensuremath{N_\mathrm{part}}\xspace}
\newcommand{\Ncoll}        {\ensuremath{N_\mathrm{coll}}\xspace}
\newcommand{\dEdx}         {\ensuremath{\textrm{d}E/\textrm{d}x}\xspace}
\newcommand{\RpPb}         {\ensuremath{R_{\rm pPb}}\xspace}

\newcommand{\nineH}        {$\sqrt{s}~=~0.9$~Te\kern-.1emV\xspace}
\newcommand{\thirteen}        {$\sqrt{s}~=~13$~Te\kern-.1emV\xspace}
\newcommand{\seven}        {$\sqrt{s}~=~7$~Te\kern-.1emV\xspace}
\newcommand{\twoH}         {$\sqrt{s}~=~0.2$~Te\kern-.1emV\xspace}
\newcommand{\twosevensix}  {$\sqrt{s}~=~2.76$~Te\kern-.1emV\xspace}
\newcommand{\five}         {$\sqrt{s}~=~5.02$~Te\kern-.1emV\xspace}
\newcommand{\twosevensixnn}{$\sqrt{s_{\mathrm{NN}}}~=~2.76$~Te\kern-.1emV\xspace}
\newcommand{\fivenn}       {$\sqrt{s_{\mathrm{NN}}}~=~5.02$~Te\kern-.1emV\xspace}
\newcommand{\LT}           {L{\'e}vy-Tsallis\xspace}
\newcommand{\GeVc}         {Ge\kern-.1emV/$c$\xspace}
\newcommand{\MeVc}         {Me\kern-.1emV/$c$\xspace}
\newcommand{\TeV}          {Te\kern-.1emV\xspace}
\newcommand{\GeV}          {Ge\kern-.1emV\xspace}
\newcommand{\MeV}          {Me\kern-.1emV\xspace}
\newcommand{\GeVmass}      {Ge\kern-.2emV/$c^2$\xspace}
\newcommand{\MeVmass}      {Me\kern-.2emV/$c^2$\xspace}
\newcommand{\lumi}         {\ensuremath{\mathcal{L}}\xspace}

\newcommand{\ITS}          {\rm{ITS}\xspace}
\newcommand{\TOF}          {\rm{TOF}\xspace}
\newcommand{\ZDC}          {\rm{ZDC}\xspace}
\newcommand{\ZDCs}         {\rm{ZDCs}\xspace}
\newcommand{\ZNA}          {\rm{ZNA}\xspace}
\newcommand{\ZNC}          {\rm{ZNC}\xspace}
\newcommand{\SPD}          {\rm{SPD}\xspace}
\newcommand{\SDD}          {\rm{SDD}\xspace}
\newcommand{\SSD}          {\rm{SSD}\xspace}
\newcommand{\TPC}          {\rm{TPC}\xspace}
\newcommand{\TRD}          {\rm{TRD}\xspace}
\newcommand{\VZERO}        {\rm{V0}\xspace}
\newcommand{\TZERO}        {\rm{T0}\xspace}
\newcommand{\VZEROA}       {\rm{V0A}\xspace}
\newcommand{\VZEROC}       {\rm{V0C}\xspace}
\newcommand{\Vdecay} 	   {\ensuremath{V^{0}}\xspace}

\newcommand{\ee}           {\ensuremath{e^{+}e^{-}}} 
\newcommand{\pip}          {\ensuremath{\pi^{+}}\xspace}
\newcommand{\pim}          {\ensuremath{\pi^{-}}\xspace}
\newcommand{\kap}          {\ensuremath{\rm{K}^{+}}\xspace}
\newcommand{\kam}          {\ensuremath{\rm{K}^{-}}\xspace}
\newcommand{\pbar}         {\ensuremath{\rm\overline{p}}\xspace}
\newcommand{\kzero}        {\ensuremath{{\rm K}^{0}_{\rm{S}}}\xspace}
\newcommand{\lmb}          {\ensuremath{\Lambda}\xspace}
\newcommand{\almb}         {\ensuremath{\overline{\Lambda}}\xspace}
\newcommand{\Om}           {\ensuremath{\Omega^-}\xspace}
\newcommand{\Mo}           {\ensuremath{\overline{\Omega}^+}\xspace}
\newcommand{\X}            {\ensuremath{\Xi^-}\xspace}
\newcommand{\Ix}           {\ensuremath{\overline{\Xi}^+}\xspace}
\newcommand{\Xis}          {\ensuremath{\Xi^{\pm}}\xspace}
\newcommand{\Oms}          {\ensuremath{\Omega^{\pm}}\xspace}
\newcommand{\degree}       {\ensuremath{^{\rm o}}\xspace}

\newcommand{\Jpsi}{\rm{J}/\psi}
\newcommand{\psit}{\psi(2{\rm S})}
\newcommand{\psip}{\psi(2{\rm S})}
\newcommand{\psipOJpsi}{\ensuremath{\psip\mbox{-over-}\Jpsi}}
\newcommand{\abs}[1]{\left\vert #1 \right\vert}
\newcommand{\ave}[1]{\langle #1 \rangle}
\newcommand{\zv}{z_{\rm vtx}}
\newcommand{\Ntr}{N_{\rm tracklet}}
\newcommand{\avNtr}{\langle N_{\rm tracklet} \rangle}
\newcommand{\Ntrc}{N_{\rm tracklet}^{\rm corr}}
\newcommand{\Ncorr}        {\ensuremath{N^\mathrm{corr}}\xspace}
\newcommand{\Ae}{A\varepsilon}
\newcommand{\mumu}{\mu^+\mu^-}
\newcommand{\BR}{\rm{BR}_{\Jpsi\rightarrow\mumu}}

\newcommand{\Ups}[1]{\Upsilon(#1{\rm S})}
\newcommand{\eightnn}       {$\sqrt{s_{\mathrm{NN}}}~=~8.16$~Te\kern-.1emV\xspace}
\newcommand{\AAcoll}          {\mbox{A--A}\xspace}

\maketitle
\section{Introduction}
\noindent Charmonium production is a complex mechanism that involves both perturbative and non-perturbative QCD aspects. Due to their heavy masses, the charm quarks  are produced during the hard process of the collision, while the formation of the bound state is a soft QCD process.
Moreover, charmonia are useful tools to investigate the medium formed in heavy-ion collisions (A-A), where a deconfined state of quarks and gluons, known as quark-gluon plasma (QGP), is expected. The study of charmonium production in smaller systems, i.e., pp and pA collisions, is essential as a reference system where no QGP effects are expected. However, several intriguing  QGP-like behaviors were observed in small systems, such as the non-zero elliptic flow of charged-particles through their long-range angular correlation measurements in high multiplicity events \cite{ALICE:2013snk}. Multiparton interactions (MPI) are one of the main scenarios proposed to explain these findings. In the MPI context, several parton-parton interactions occur in the same hadron collision. The multiplicity dependence of charmonium production is an indirect probe for MPI. It provides information on the interplay between the soft and hard QCD process, as heavy quark production is a hard process while charged particles are produced via soft QCD.
In these proceedings, the multiplicity dependence of the self-normalised $\psip$ yields in pp and p--Pb collisions at $\s$ = 13 TeV and $\snn$ = 8.16 TeV, respectively, are discussed. The $\psip$-over-J/$\psi$ ratio as a function of charged-particle multiplicity is also shown. \\    

\section{Experimental setup}
\noindent A detailed description of the ALICE detector and its performance can be found in \cite{ALICE:2014sbx}. The analyzed data are recorded using the muon spectrometer, located in the forward arm of the ALICE detector and  covering the rapidity window of 2.5 < $\eta\rm{_{lab}}$ < 4.0. The muon spectrometer consists of the following subdetectors: (i) a set of absorbers that are essential to reduce the background contamination coming from the decay of hadrons and beam-gas interaction, (ii) a dipole magnet that deflects particle trajectories to measure their momentum and electric charge, (iii) five stations of muon tracking chambers to reconstruct the particle trajectory and two stations of muon trigger chambers which are essential for the muon identification.      
Other detectors are also used in this analysis, such as the silicon pixel detector (SPD) and the V0 detector. The SPD (|$\eta$| < 2) is important for the reconstruction of the charged-particle tracks and the primary vertex. The V0 detector, which consists of two sets of scintillator arrays located at backward -3.7 < $\eta$ <-1.7  and forward 2.8 < $\eta$ < 5.1 rapidity, serves as a minimum bias trigger in this analysis. 
\section{Results}
\begin{figure}[!htbp]
 \centering
    \subfigure[$\psi(2S)$ self-normalised yield.]{
    \includegraphics[width =0.45\textwidth]{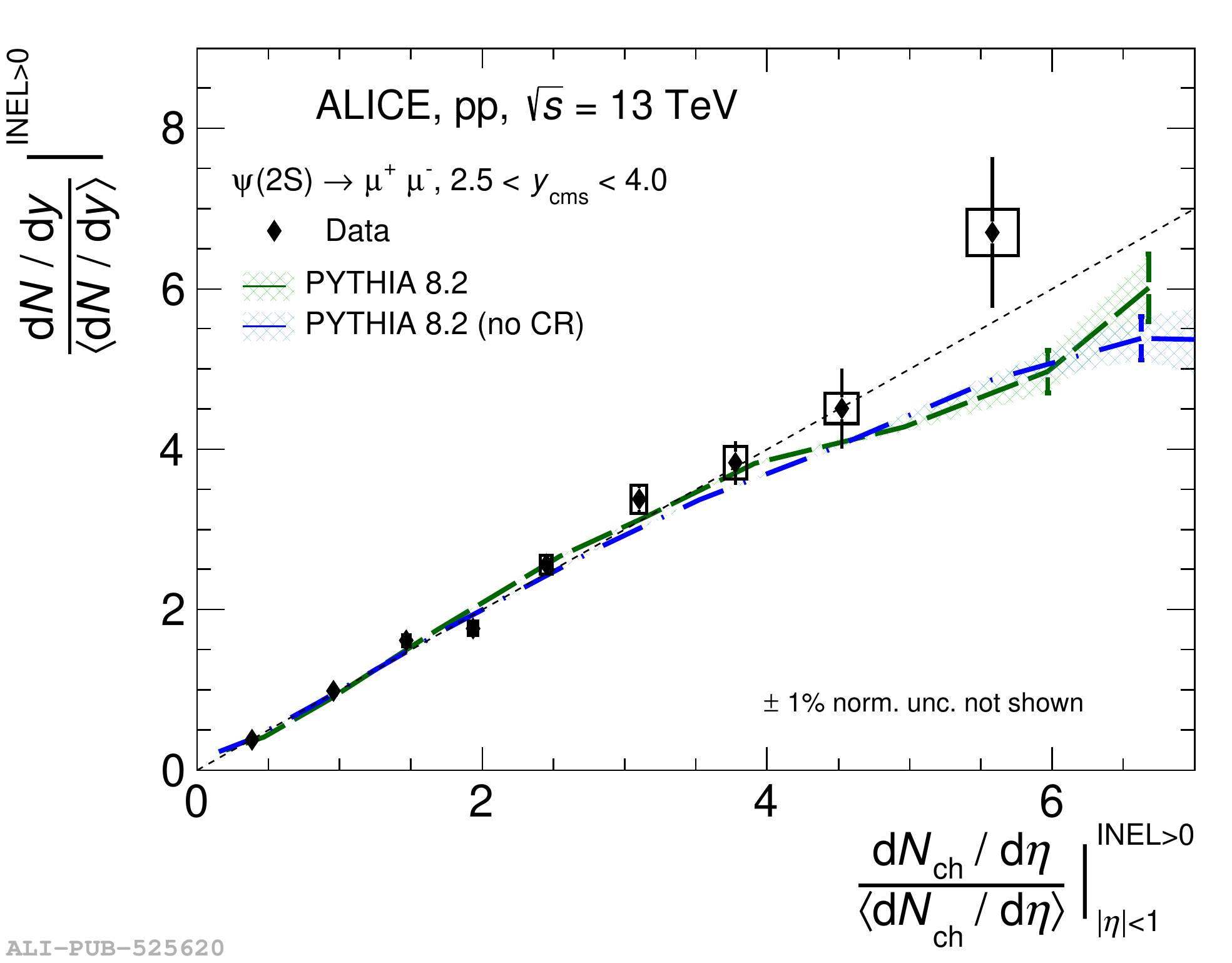}
    \label{fig:Psi2SPPModels}
    }
    \qquad
    \subfigure[$\psi(2S)$-to-J/$\psi$ self-normalised yield ratio.]{
    \includegraphics[width =0.45\textwidth]{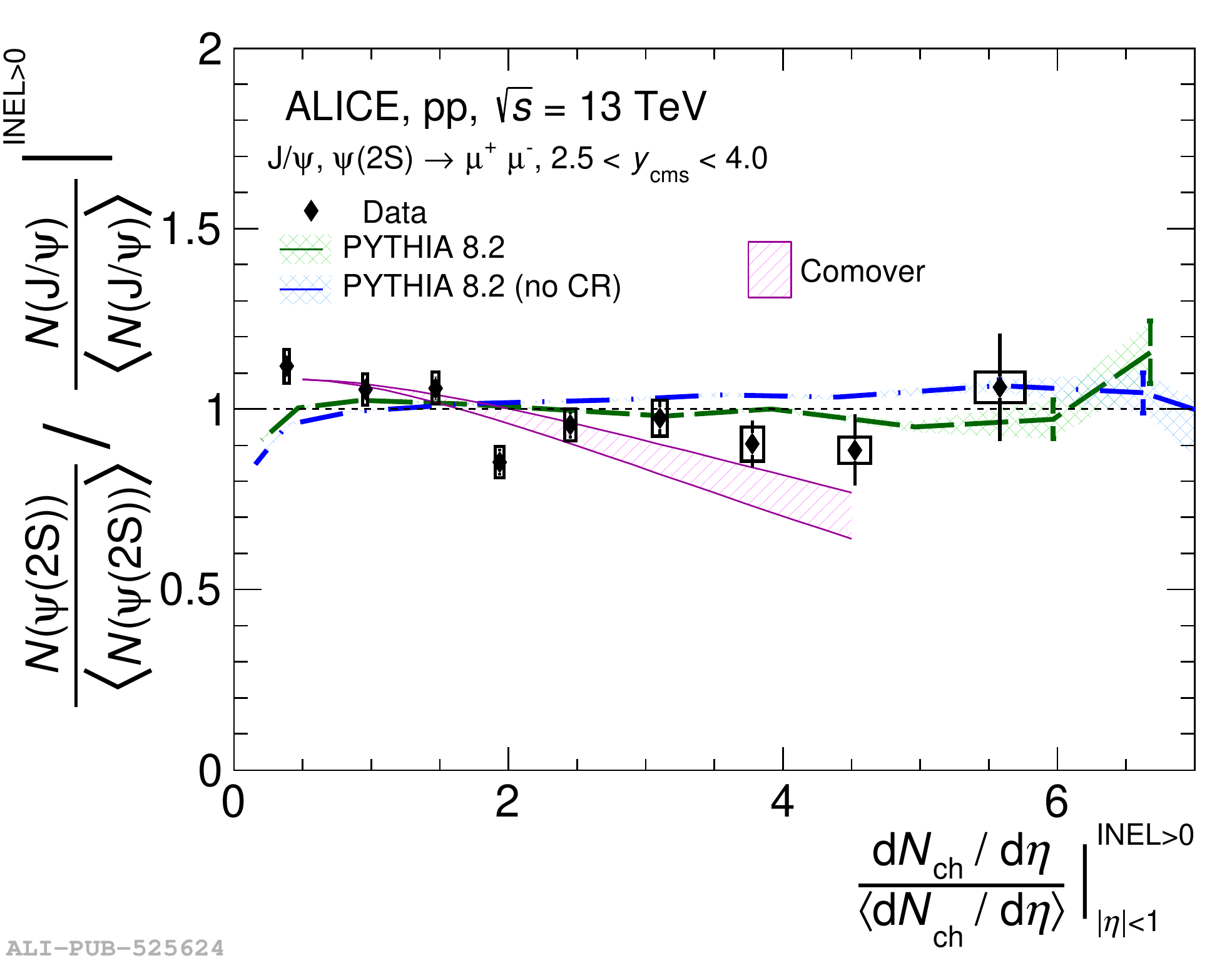}
    \label{fig:psi2sOverJPsiPP}
  }
  \caption{ (a) self-normalised $\psi(2S)$ yields, measured at forward rapidity, as a function of the midrapidity charged-particle multiplicity, in pp collisions at $\s$ = 13 TeV \cite{ALICE:2022gpu}. (b) $\psi(2S)$-to-J/$\psi$ self-normalised yield ratios, measured at forward rapidity, as a function of the  midrapidity charged-particle multiplicity \cite{ALICE:2022gpu}. The results are compared to PYTHIA 8.2 calculations and comovers model \cite{Sjostrand:2014zea,Ferreiro:2014bia}. }
\end{figure}

\noindent The multiplicity dependence of the self-normalised $\psip$ yield in pp collisions at \mbox{$\s$ = 13 TeV} is reported in \cite{ALICE:2022gpu}. In this analysis, the $\psip$ mesons are reconstructed from their decay channel into dimuons in the rapidity window 2.5 < $y\rm{_{lab}}$ < 4.0, while the charged-particle multiplicity is measured at midrapidity |$\eta\rm{_{lab}}$| < 1. The $\psip$ yields and the charged-particle multiplicity are normalised to their respective average values evaluated in the integrated multiplicity bin. As shown in Fig.\ref{fig:Psi2SPPModels}, the $\psip$ normalised yields increase with charged-particle multiplicity within uncertainties. The resulting trend is compatible with a linear increase depicted with a dashed line in the figure. The result is compared to PYTHIA 8.2 calculations \cite{Sjostrand:2014zea}, with and without color reconnection (CR). PYTHIA 8.2 event generator describes the production of the charm quark during the initial hard processes and with MPI. The number of charged particles produced is proportional to the number of events containing MPI. In  PYTHIA 8.2, the color reconnection scenario represents the fusion of the final state partons produced via MPI. In Fig.\ref{fig:Psi2SPPModels} PYTHIA 8.2 calculations, both with and without color reconnection, show similar behavior for the $\psip$ production as a function of multiplicity. The trend of the calculation is compatible with measurements up to 5 times the average multiplicity.      

\noindent The $\psi(2S)$-over-J/$\psi$ self-normalised yield ratio as a function of the midrapidity charged-particle multiplicity is presented in Fig.\ref{fig:psi2sOverJPsiPP}. The double ratio measurement allows one to disentangle any possible differences in the multiplicity dependence between the charmonium excited and ground states. The resulting trend is compatible with a flat behavior, within uncertainties. The result is compared to PYTHIA 8.2 event generator calculations which predict a similar behavior for both particles within uncertainties. The result is also compared to the comovers model calculation \cite{Ferreiro:2014bia}. In the comovers model, charmonia interact with the final state particles that are comoving with them. The dissociation probability depends on the size of charmonioum state and the density of comoving particles. The comovers model predicts a stronger suppression for the $\psi(2S)$ with respect to J/$\psi$, as the former has a larger size than the J/$\psi$ \cite{Ferreiro:2012fb}. 
The multiplicity dependence of the J/$\psi$ production in pp collisions at $\s$ = 13 TeV can be found in \cite{ALICE:2021zkd}.

\begin{figure}[!htbp]
 \centering
    \subfigure[$\psi(2S)$ self-normalised yield in pp and p--Pb collisions.]{
    \includegraphics[width =0.45\textwidth]{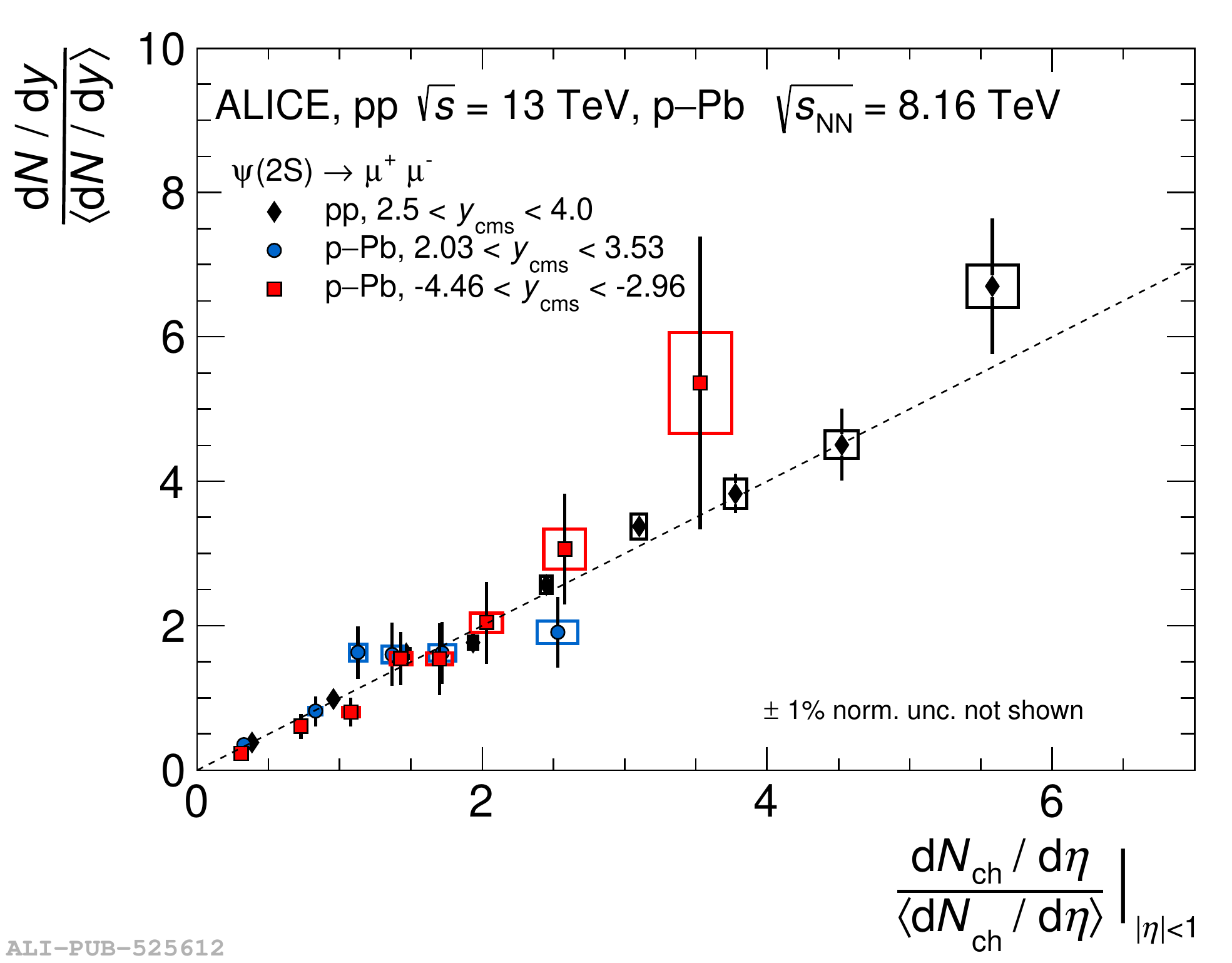}
    \label{fig:Psi2SPPPBALL}
    }
    \qquad
    \subfigure[$\psi(2S)$ self-normalised yield Pb-going direction.]{
    \includegraphics[width =0.45\textwidth]{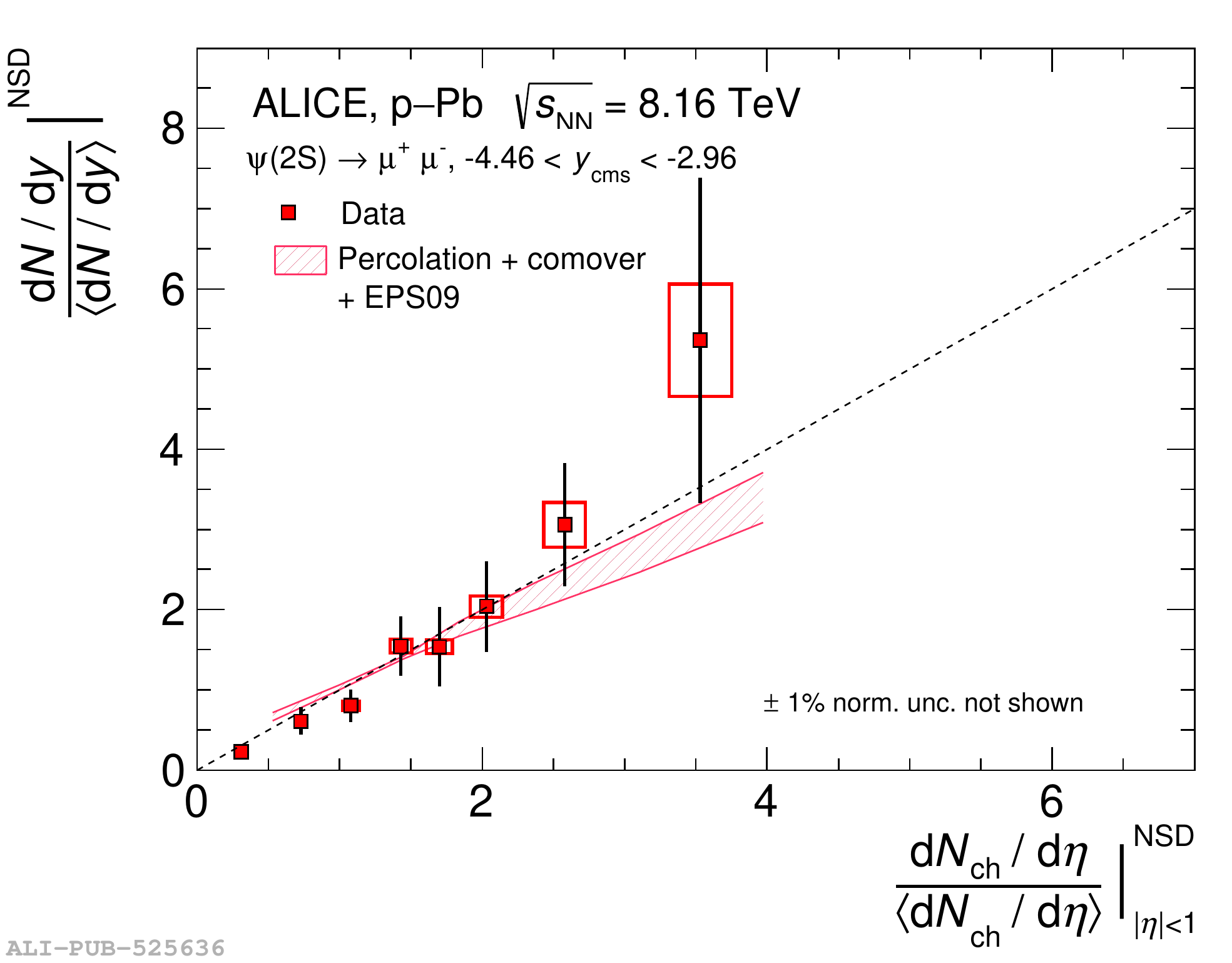}
    \label{fig:psi2PbBW}
  }
  \caption{(a) $\psip$ self-normalised yields as a function of the charged-particle multiplicity in pp collisions at $\s$ = 13 TeV and in p--Pb collisions at $\snn$ = 8.16 TeV, for both p- and Pb-going direction. (b) $\psip$ self-normalised yields as a function of the charged-particle multiplicity in p--Pb collisions, for the Pb-going direction, at $\snn$ = 8.16 TeV \cite{ALICE:2022gpu}. The result is compared to the Percolation + comover + EPS09 model   \cite{Ferreiro:2014bia,Ferreiro:2012fb,Drescher:2000ha}. }
 \end{figure}

\noindent From the theoretical point of view, the production of hadrons in pA collisions is expected to be affected by several nuclear effects, so called cold nuclear matter effects.  For example, the gluon shadowing \cite{Vogt:2004dh} is responsible for the suppression of charmonia in p--Pb collisions with respect to pp collisions. This is due to the modification of the nuclear parton distribution functions in the nucleus compared to free nucleons. The percolation model \cite{Ferreiro:2012fb} represents the initial state partons as strings that have a specific size. These strings can interact and overlap at sufficiently high energies, reducing the total effective number of strings and suppressing the charmonioum production. The effects above are expected to affect equally both J/$\psi$ and $\psi(2S)$ production, as they are sensitive to the initial state partons.             
\noindent The multiplicity dependence of the $\psip$ production in p--Pb collisions at $\snn$ = 8.16 TeV is presented in \cite{ALICE:2022gpu}. The dataset used in this analysis was recorded with two different beam configurations. The first one is obtained when the proton beam is directed from the ALICE interaction point (IP) toward the muon spectrometer, the so-called p-going direction. In this case, the charmonium yields are reconstructed in the center of mass rapidity window 2.03 < $y\rm{_{cms}}$ < 3.53.  The second configuration is obtained by inverting the direction of the colliding beams, i.e. the direction of the Pb beam is from IP toward the muon spectrometer (Pb-going direction). The charmonium yields in this case are reconstructed in the backward center of mass rapidity -4.46 < $y\rm{_{cms}}$ < -2.96. Fig.\ref{fig:Psi2SPPPBALL} shows the multiplicity dependence of $\psip$ yields in p--Pb collisions for both p-going and Pb-going directions compared to the result from pp collisions. The $\psip$ yield shows an increasing trend with increasing charged-particle multiplicity for both configurations. The trend of the result is compatible with the measurements from pp collisions within uncertainties. Fig.\ref{fig:psi2PbBW} presents the $\psip$ self-normalised yields as a function of charged-particle multiplicity compared to the percolation coupled to comover model and using the EPS09 nuclear parton distribution function (nPDF). The Percolation + comover + EPS09 model is compatible with the result within uncertainties. The self-normalised $\psi(2S)$-over-J/$\psi$ yield ratio, shown in \cite{ALICE:2022gpu}, is compatible with a flat distribution within the large experimental uncertainties for both backward and forward rapidity in p--Pb collisions. The comovers model suggests a stronger suppression for the $\psi(2S)$ with respect to the J/$\psi$, especially at high multiplicities (see \cite{ALICE:2020eji} for the J/$\psi$ multiplicity dependence in p--Pb collisions at $\snn$ = 8.16 TeV). A stronger suppression is expected at backward rapidity as compared to forward rapidity due to the larger density of soft particles, which increases the probability of interaction with charmonia and, therefore, the dissociation probability. Within the large experimental uncertainties, the comover predictions are compatible with the data (although there is no evidence for a decreasing trend with multiplicity of the $\psi(2S)$-over-J/$\psi$ ratio).
\section{Conclusion}
\noindent The $\psip$ self-normalised yields as a function of the charged-particle multiplicity in pp and p--Pb collisions at $\s$ = 13 TeV and $\snn$ = 8.16 TeV, respectively, are discussed. The results show an increase in the self-normalised yields as a function of multiplicity for both data sets. This increase is compatible with a linear trend within uncertainties. The trend of the result is described by several calculations, which model initial state effects (including MPI) and/or final state effects. The multiplicity dependence of $\psi(2S)$-over-J/$\psi$ self-normalised yield ratio is also reported. The trend of the double ratio is flat within uncertainties, suggesting a similar behavior for the ground and excited charmonium states as a function of multiplicity. The LHC Run 3 will provide a data set with higher statistics allowing for significantly more precise measurements with an increased multiplicity reach, therefore adding more constraints on the MPI modeling.
{
\small
\bibliographystyle{utphys}   
\bibliography{alice_pub,models}
}
\end{document}